\documentclass[sigconf]{acmart}
\AtBeginDocument{%
  \providecommand\BibTeX{{%
    \normalfont B\kern-0.5em{\scshape i\kern-0.25em b}\kern-0.8em\TeX}}}

\setcopyright{acmcopyright}
\copyrightyear{2021} 
\acmYear{2021} 
\setcopyright{acmlicensed}\acmConference[AIES '21]{Proceedings of the 2021 AAAI/ACM Conference on AI, Ethics, and Society}{May 19--21, 2021}{Virtual Event, USA}
\acmBooktitle{Proceedings of the 2021 AAAI/ACM Conference on AI, Ethics, and Society (AIES '21), May 19--21, 2021, Virtual Event, USA}
\acmPrice{15.00}
\acmDOI{10.1145/3461702.3462590}
\acmISBN{978-1-4503-8473-5/21/05}

\usepackage{booktabs}
\usepackage{arydshln}

\usepackage{url}

\begin{document}

\title{Understanding the Representation and Representativeness of Age in AI Data Sets}

\author{Joon Sung Park}
\affiliation{%
  \institution{Stanford University}
  \city{Stanford}
  \state{CA}
  \country{USA}}
\email{joonspk@stanford.edu}

\author{Michael S. Bernstein}
\affiliation{%
  \institution{Stanford University}
  \city{Stanford}
  \state{CA}
  \country{USA}}
\email{msb@cs.stanford.edu}

\author{Robin N. Brewer}
\affiliation{%
  \institution{University of Michigan}
  \city{Ann Arbor}
  \state{MI}
  \country{USA}}
\email{rnbrew@umich.edu}

\author{Ece Kamar}
\affiliation{%
 \institution{Microsoft Research -- Redmond}
 \city{Redmond}
 \state{WA}
 \country{USA}}
\email{eckamar@microsoft.com}

\author{Meredith Ringel Morris}
\affiliation{%
 \institution{Microsoft Research -- Redmond}
 \city{Redmond}
 \state{WA}
 \country{USA}}
\email{merrie@microsoft.com}

\sloppy

\renewcommand{\shortauthors}{Joon Sung Park, et al.}
\newcommand\revision[1]{\textcolor{red}{#1}}

\begin{abstract}
  A diverse representation of different demographic groups in AI training data sets is important in ensuring that the models will work for a large range of users. To this end, recent efforts in AI fairness and inclusion have advocated for creating AI data sets that are well-balanced across race, gender, socioeconomic status, and disability status. In this paper, we contribute to this line of work by focusing on the representation of age by asking whether older adults are represented proportionally to the population at large in AI data sets. We examine publicly-available information about 92 face data sets to understand how they codify age as a case study to investigate how the subjects’ ages are recorded and whether older generations are represented. We find that older adults are very under-represented; five data sets in the study that explicitly documented the closed age intervals of their subjects included older adults (defined as older than 65 years), while only one included oldest-old adults (defined as older than 85 years). Additionally, we find that only 24 of the data sets include any age-related information in their documentation or metadata, and that there is no consistent method followed across these data sets to collect and record the subjects’ ages. We recognize the unique difficulties in creating representative data sets in terms of age, but raise it as an important dimension that researchers and engineers interested in inclusive AI should consider. 
\end{abstract}




\maketitle

\section{Introduction}
Our society is getting older. Today, more than 15\% of the U.S. population is 65 years old or older \cite{ACL2017}, and by 2050 this proportion will be matched globally \cite{UN2019}. Additionally, studies suggest the ``oldest-old'' population, defined as those who are at least 85 years old \cite{Campion1994, Cohen2013, Loi2021}, will see the greatest rate of increase \cite{Pollack2005}. This will mark a significant change in the makeup of our society, and it will become increasingly more important to ensure that emerging AI-infused systems are inclusive of this large and growing population who may benefit from the power that AI brings to both general-purpose and health-related domains.  

A key component for creating inclusive AI systems that work for a diverse group of users is ensuring representation of diverse popluations in the data used to train and test ML models. A growing number of evaluations have explored AI systems’ performance disparities for people with marginalized demographic attributes that often originate from biased data sets used to train them \cite{3_Whittaker2019}. These works have investigated how commonly-used AI systems such as facial analysis or speech recognition fail to achieve the same level of performance on the basis of gender \cite{Buolamwini2018, Lohr2018, Rodger2004}, race \cite{Buolamwini2018, Lohr2018}, socioeconomic status \cite{FB2019}, and disability status \cite{33_Guo2020, 3_Whittaker2019}, but found that such disparities can often be mitigated by updating the model using  using a more balanced data set across different demographics \cite{18_Puri2018}. In this work, we extend this line of effort to discuss whether the AI data sets used today represent the older adult population, a group that has been subjected to negative societal attitudes and stereotypes in the form of ageism \cite{Butler1969}.

In particular, we use facial analysis systems as a case study for observing how the older adult population is represented in the data sets that are used to train such AI systems. We focus on facial analysis systems as they are particularly relevant for understanding how our identity is operationalized in today's AI systems \cite{Scheuerman2020}, but we suggest that our work is the first step towards investigating age-related AI performance that should be expanded into other areas of AI. We started our work by drawing from a list of 92 face image data sets based on 277 academic publications that the authors of a prior work compiled to study how people’s genders were classified in the training data of facial analysis systems  \cite{Scheuerman2020}. Using the publicly available documentation of these data sets (n=92), we analyze how and why the data was collected, particularly focusing on the presence of age-relevant metadata for the subjects and the data sets' coverage of the older adults' age bracket. Additionally, we selected 31 of the data sets in the list that are still publicly downloadable with clear terms of service that complied with our institution’s IRB to further inspect how age was actually codified in these data sets. Specifically, we ask the following research questions: 

\begin{enumerate}
  \item Is the age-related information of the subjects included in the metadata of the data set or its documentation? If so, how was the age binned (i.e. did the data set include the specific age of the subject or were numerical brackets or broad age descriptors used, and if so, and how were such brackets defined)? 
  \item What was the process for annotating the subjects’ age (i.e. was it directly sourced by the subjects when their photos were taken or was it derived afterward such as by third-party labelers)?
  \item What was the goal of creating the data set, and how did this interact with whether age was included in the metadata (i.e. was age included to train algorithms for age-related classification tasks like age estimation)? 
  \item Is the older adult population (aged 65+) represented proportionally to the population at large? Does this representation, or lack thereof, extend to the oldest-old adults population (aged 85+)?  
\end{enumerate}

We find that only 26\% of the 92 face image data sets and their documentation contain any age-related metadata about their subjects. Furthermore, for these data sets, the norms of how to report the subjects’ ages are highly inconsistent, with some data sets simply documenting age in a binary category of “young” and “old,'' while others using unevenly spaced brackets where the older age brackets encompass a much wider range than younger brackets. In addition, the age metadata in these data sets rarely acknowledges older adults, with the highest age bracket often ending with 50 years old or older or even lower. The few exceptions to this were specialized data sets that were collected to train algorithms for age estimation, but even in these data sets the age distribution includes only a small number of older adults and very few (or zero) oldest old adults. Finally, we find that even in those data sets where age was included, this information often was not verified nor sourced directly from the subjects but instead was annotated by crowdworkers who guessed the subjects’ age based on their appearance or inferred it using publicly  available information (such as the date of birth for celebrities).

The contribution we make in this work is focused, but poignant; our findings suggest that the representation of older adults aged 65+ in popular data sets used to train AI systems for facial analysis is severely lacking, while that of the oldest-old adults aged 85+ is almost none. This lack of representation is, though not necessarily surprising, more severe than what one might expect -- only five out of 92 data sets explicitly included an age bracket that covers older adults and only one included an age bracket that covers oldest-old adults. Taking the face data sets as one example of lack of representation for older adults, there is cause for concern as to whether other classes of AI training data are representative of diverse ages and whether newer AI-infused technologies will work well for this fast-growing population.  Given this, we highlight the need for better representation of the older adults in AI data sets, and the need for standardized procedures for documenting age metadata. 

\section{Related Work}
We summarize prior work that investigated how older adults may benefit from AI technology. We then cover the growing concern around biases and performance disparities that AI systems exhibit in relation to users' demographic traits and how the lack of representation of certain groups of people in training data sets can aggravate such outcomes.

\subsection{AI Bias and Under-Representation}
An important on-going challenge in AI and ethics has been that of bias and performance disparities of AI systems for people with historically marginalized demographic attributes. A growing number of studies have shown that one’s race \cite{Buolamwini2018, Lohr2018}, gender \cite{8_Nicol2002, 10_Tatman2017, Scheuerman2020}, socioeconomic status \cite{FB2019}, and disability status \cite{3_Whittaker2019, Park2021, 1_Guo2020} can lower the performance of AI systems like facial recognition or natural language processing systems. The source of this challenge has often been attributed to the under-representation of marginalized populations in AI training data sets as the models learn to perceive the world based on what they are given, and if a certain demographic population is missing, they will inevitably fail to recognize that population \cite{3_Whittaker2019}. For instance, the Gender Shades study showed that commercial AI systems that are used for binary gender classification based on one’s appearance often fail for women of darker skin color \cite{Buolamwini2018}. Following this, the developers of such systems updated their models with a more balanced training data set to remedy these shortcomings, reducing the error rate by nearly ten-fold when tested against a similar data set to what was used in the Gender Shades study \cite{18_Puri2018}. As a response to this, there have been calls for action and efforts to create more balanced data sets in terms of race, gender, socioeconomic status, and disability status. However, the dimension of age, which is the focus of our work, has received very little attention in the context of AI data representation despite its importance.

\subsection{Aging Population and Technology}
The global population is aging as life expectancy rises. The United Nations reports that this trend, which first emerged among developed countries, is now observed in virtually all developing countries \cite{77_UN}. Globally, a large increase is expected among older adults (defined as those at least 65 years of age), a group that is expected to nearly double by 2050 \cite{UN2019}, while in the U.S., a particularly significant increase is expected among the oldest-old adults (defined as those at least 85 years of age), a group that is expected to represent 4.3\% of the nation’s population by 2050 \cite{AARP2010}. 

While we expect that many in the older age bracket will remain healthy and productive, many will also experience physical and cognitive impairment at a higher rate than those who are younger \cite{Hebert2003}. Physical impairment \cite{Berkman1993}, age-related mobility disability due to decreased strength \cite{Guralnik2000}, and sensory deficits \cite{Berkman1993, Jeste2013} commonly accompany ageing, and the rate of clinical depression and loneliness rises as older adults grieve losses and their social ties decline \cite{Adams2004}.

AI-infused assistive technologies and other AI tools present opportunities to significantly support the needs of older adults and help them age on their own terms \cite{Jeste2013, Zuckerman2020, Faber2001, Montross2006}. As older adults increasingly prefer to age-in-place or live in their homes rather than long-term care \cite{AARP2018}, AI can offer them greater mobility through smart navigation or robotics \cite{NSF2018}, control over their living spaces for more independence through smart homes \cite{Trajkova2020, Koshy2021}, and access to on-demand medical expertise through powerful medical recommender systems \cite{Angulu2018}. AI can also be used to augment existing technologies and online communities that older adults use. For example, facial recognition systems for unlocking phones can remove or reduce barriers for older people with limited motor control and/or experiencing cognitive decline so they do not need to type on a small phone keyboard or remember passwords \cite{xfinity2019}. AI can also support natural language interfaces, removing barriers to keyboard or keypad input \cite{Trajkova2020, Rieland2017}, thereby making technologies easier and more natural to use even for those older adults with low computer literacy. Such advances can not only allow older adults to use digital communication to engage with their social networks \cite{Angelini2016, Ballagas2010, Davis2007}, but also combat the stereotypical premise that older adults lack the desire to use technologies \cite{Loi2021, Coleman2010}.

However, without proper representation of older adults in data sets used to train and test AI models, it is difficult to ensure that the new generation of AI technology will work for this population. Here, we highlight that age is a demographic category that is difficult to balance but potentially highly impactful and worth considering for our research community. In the remainder of the paper, we take the first step towards considering age representation in AI data sets by studying 92 face data sets that are used to train facial recognition and analysis systems to see how age is represented in these data sets. Our focus on face data sets is motivated in large part by the fact that facial analysis technologies are “particularly pertinent to understanding how identity is operationalized in new technical systems” \cite{Scheuerman2020}. But our study on face data sets is also a case study that should inspire similar explorations on other forms of AI data sets. 
\section{Method}
Our aim is to conduct a broad investigation of how face image data sets represented age-relevant metadata for their subjects by analyzing the data sets and the documentation offered in their relevant academic publications. To this end, we take advantage of a recently compiled list of face image data sets referenced in research publications and conduct analysis on those data sets and publications. In this section, we briefly summarize how the list was compiled and explain our methods of analysis. 

\subsection{Collecting the Data Sets}
In a recent 2020 study, Scheuerman et al. investigated how people’s gender and race were codified into face image data sets by gathering an extensive list of such data sets that were  published by academic researchers \cite{Scheuerman2020}. This list was made public as a part of the 2020 study.\footnote{The list is available for download through the following DOI: 10.5281/zenodo.3735400} We used the data sets included in this list as the basis for our study. Scheuerman et al. generated this list by taking the following approach \cite{Scheuerman2020}: 

They first gathered a corpus of research papers that were published by two of the largest associations for computing research, the Association for Computing Machinery (ACM) and the Institute of Electrical and Electronics Engineers (IEEE). This was done by scraping or downloading manuscripts from the respective communities’ digital archive. For the papers published in the ACM,  Scheuerman et al. scraped 18,661 manuscripts from ACM Digital Library’s (ACM DL) search results for “facial recognition” and for those published in the IEEE,  Scheuerman et al. used IEEE’s Xplore library to export 4,000 manuscripts using the search terms ``facial recognition” and “facial classification.” This corpus was then narrowed down by filtering the author-provided keywords for “facial recognition,” “face recognition,” “face classification,” and finally with a publication period that ranged from 2014 to 2019. This resulted in 277 manuscripts from which a final list of data sets included 92 image data sets that had publicly available documentation. 

Given this list \cite{Scheuerman2020}, we further identified the data sets that are still publicly downloadable (as of October 2020) and have terms of service. For the 31 data sets that were still available in October 2020 and also complied with our institution’s IRB data set onboarding process, we proceeded to download the data sets in order to inspect how age information is represented in their metadata. For the remaining 61 data sets, we only analyze the documentation included in the original academic publications that introduced the data set or the main download pages of the data sets that illustrate the contents of the data set and how it was collected. We report findings from our analysis of the documentation of all 92 image data sets, and use our findings from inspecting the metadata of the 31 data sets we were able to download to illustrate the trends we find.

\begin{table*}[th]
\centering
\caption{The categories of age-related information reported in the data set documentation; only 24 of the 92 data sets contained any age-related documentation. The asterisk denotes data sets we were also able to download and inspect per their ongoing availability and terms of service. We report the maximum age and age distribution in the cases where we could determine this information from the data set or its documentation.}
\begin{tabular}{llcl}
\textbf{Face Data Set}                       & \textbf{Age-Related Information}                                                              & \multicolumn{1}{l}{\textbf{Max Age}} & \textbf{Age Distribution}                                                                \\ 
\hline\hline
Computer Vision Laboratory (1999)            & Approximation (ie. around 18)                                                                 & Unknown                              & Unknown                                                                                  \\ 
\hdashline[1pt/1pt]
Cohn-Kanade (2000)                           & Age range                                                                                     & 50 years                             & \%65+ =0\%                                                                               \\ 
\hdashline[1pt/1pt]
Gavab (2004)                                 & Age range                                                                                     & 40 years                             & \%65+ =0\%                                                                               \\ 
\hdashline[1pt/1pt]
The IMM Frontal Face (2005) *                & Raw age                                                                                       & Unknown                              & \begin{tabular}[c]{@{}l@{}}mean=31.9 (std=8.4);\\\%65+ =0\%\end{tabular}                 \\ 
\hdashline[1pt/1pt]
MORPH (2006)                                 & Age range, age distribution                                                                   & 77 years                             & Unknown                                                                                  \\ 
\hdashline[1pt/1pt]
Iranian Face (2007)                          & Binned categories                                                                             & 85 years                             & \begin{tabular}[c]{@{}l@{}}\%61+ =7.47\%; \\\%71+ =3.73\%; \\\%81+ =0.81\%\end{tabular}  \\ 
\hdashline[1pt/1pt]
CAS-PEAL (2007)                              & Binned categories                                                                             & 74 years                             & Unknown                                                                                  \\ 
\hdashline[1pt/1pt]
NimStim Set of Facial Expressions (2009)     & Age distribution                                                                              & 35 years                             & mean=25.8 (std=4.1)                                                                      \\ 
\hdashline[1pt/1pt]
MUCT (2010) *                                & Minimum age                                                                                   & Unknown                              & Unknown                                                                                  \\ 
\hdashline[1pt/1pt]
Cohn-Kanade + (2010)                         & Age range                                                                                     & 50 years                             & \%65+ =0\%                                                                               \\ 
\hdashline[1pt/1pt]
NVIE (2010)                                  & Age range                                                                                     & 31 years                             & \%65+ =0\%                                                                               \\ 
\hdashline[1pt/1pt]
Radboud Faces Database (2010)                & Age distribution                                                                              & Unknown                              & mean=21.2 (std=4.0)                                                                      \\ 
\hdashline[1pt/1pt]
CMU Multi-PIE Face (2010)                    & \begin{tabular}[c]{@{}l@{}}Birth year \& approx. image taken,\\age distribution \end{tabular} & Unknown                              & Unknown                                                                                  \\ 
\hdashline[1pt/1pt]
CIFAR-100 (2011) *                           & Abstract age categories                                                                       & Unknown                              & Unknown                                                                                  \\ 
\hdashline[1pt/1pt]
Static Facial Expressions in the Wild (2011) & Birth year  approx. image taken                                                               & 70 years                             & Unknown                                                                                  \\ 
\hdashline[1pt/1pt]
Indian Movie Face (2013) *                   & Binned categories                                                                             & Unknown                              & Unknown                                                                                  \\ 
\hdashline[1pt/1pt]
10k US Adult Faces (2013)                    & Binned categories                                                                             & Unknown                              & median=30-40~                                                                            \\ 
\hdashline[1pt/1pt]
Long Distance Heterogeneous Face (2014)      & Age range                                                                                     & 30 years                             & \%65+ =0\%                                                                               \\ 
\hdashline[1pt/1pt]
Cross-Age Celebrity (2014) *                 & Birth year  approx. image taken                                                               & 62 years                             & \%65+ =0\%                                                                               \\ 
\hdashline[1pt/1pt]
GUC Light Field Face Artifact (2015)         & Binned categories, age distribution                                                           & Unknown                              & \%31+ =15.0\%                                                                            \\ 
\hdashline[1pt/1pt]
Tri-Subject Kinship Verification (2015)      & Kinship relationship                                                                          & Unknown                              & Unknown                                                                                  \\ 
\hdashline[1pt/1pt]
Microsoft Celeb (2016)                       & Birth year  approx. image taken                                                               & Unknown                              & Unknown                                                                                  \\ 
\hdashline[1pt/1pt]
Large Age-Gap (2017) *                       & Abstract age categories                                                                       & Unknown                              & Unknown                                                                                  \\ 
\hdashline[1pt/1pt]
Real-world Affective Faces Database (2017)   & Age range                                                                                     & 70 years                             & Unknown                                                                                  \\
\bottomrule
\end{tabular}
\end{table*}

\subsection{Analyzing the Data Sets}
By studying the data sets and their documentation, we aimed to find out whether 1) age-related information about the subjects is included in the data sets or in their documentation, 2) older adults are represented proportionally to the population at large, 3) the goal of the data set interacted with whether and how age was codified, and 4) how age of the subjects was annotated. In order to quantify our observations, we iteratively developed a codebook to codify our data, as described below:   

\subsubsection{Age-related information} To summarize whether age-related information is included in the data sets, we coded a data set with “present” if it includes any information about a person's age. This included those data sets that documented either the age distribution or the raw age for their subjects.

\subsubsection{Older adult representation by bracket} Prior literature that connects age and technology notes that "older adult" is a broad term used to categorize those of age 65 and older and that it can be subdivided further: the youngest old (65-74), the middle old (75–84), and the oldest old (85+) \cite{Campion1994, Cohen2013, Loi2021}. We used these three age categories in our coding. 

\subsubsection{The goal of a data set} Scheuerman et al.’s findings noted that the face data sets in the list had three broad categories of use cases \cite{Scheuerman2020}: 1)~for individual face recognition or verification, 2)~for image labeling or classification, and 3)~for adding diversity to training and evaluation data. In this work, we are interested in understanding an additional goal of these data sets, i.e., whether the creators anticipated any uses or issues with respect to age that motivated the inclusion of age-related information in the metadata. We coded  data sets with any explanations their documentation provided for including the subjects’ age and recorded the emergent themes that arose. 

\subsubsection{Age annotation scheme} Finally, among the data sets that contained some form of age-related information, we coded how the age of the subjects was annotated. Our main focus was to distinguish the following three types of annotation schemes: 1)~recording the actual age of the subjects provided by the subjects themselves, 2)~inferring the subjects’ age using other metadata, and 3)~estimating the subjects’ age by observing their appearance. 
\section{Results}

% \msb{Is there a page limit to this submission? If not, or if we're not abutting that page limit, I think there is an opportunity for a particularly compelling figure here. In particular, take the oldest face in each of the datasets (or a sample from the oldest bracket they have) and array them all out into a tiled view. By your analysis saying that the oldest individual was 56yo, they should all look very young in that tiling.}

In this section, we present our findings, organized by our research questions presented at the beginning of this paper. 

\subsection{Age-Related Information}
We find that a majority of the face image data sets in our study did not include any age-related information. Of the 92 face data sets whose documentation we studied, only 24 mentioned some form of collecting age-related information of their subjects (26\%). Similarly, of the 31 data set downloads that we studied, only 6 of the data sets contained age-related metadata (19\%). It is worth noting that all data sets with such metadata also included information about their subjects' age in their documentation. 

Of those that included age information, only 20\% of the documentation and 33\% of the data sets included or mentioned the subjects’ raw age or date/year of birth. The rest codified the subjects’ age in some aggregate forms but without any consistent standards across them. Among the documentation of data sets that contained age-related information, two most common way of aggregating the subjects' ages was to simply provide an overall range for the ages included in the data set (29\%) and to bin by age group categories with the range of each age group numerically defined (25\%). However, the age groups used to report the subjects’ age were inconsistent both in terms of the range of each age category and the starting and ending age for the age field. For example, the 10K U.S. Adult Faces Database that contains over ten thousand images of  U.S. adults used age group categories that included 20-30, 30-45, 45-60, and over 60 years old \cite{66_Bainbridge}, whereas CAS-PEAL with images of 1,040 individuals used categories that included 18-44, 45-59 and 60-74 years old \cite{67_Gao}. Meanwhile, some documentation simply reported the age distribution of the subjects, for instance in the form of the average and standard deviation (21\%), or simply noted the subjects’ age requirement for participation (4\%). 

Of the 6 data sets that we were able to download which also had age-related metadata, we found that 33\% of them used more abstract categories to describe age that illustrate the rough approximation of the subjects’ age but that are only weakly defined and open to the interpretation of annotators. For example, the Large-Age Gap data set that contains images of 1,010 celebrities, each with images from when they were young and old, simply codified the subjects’ age with a binary field of “young” and “old” in its metadata without numerical definitions \cite{69_Bianco}. Similarly, CIFAR-100, which contains a large number of annotated images, included categories of “baby,” “boy,” “girl,” “man,” and “woman” without providing precise definitions that distinguish the age of a boy from man and a girl from woman \cite{68_Krizhevsky}.

\subsection{Annotating Subjects’ Age}
% \msb{when doing an inline enumeration like this, use a nonbreaking space after each number (1)~foo. That way we don't have the number cut off from its text on the next line}
We found that the method for annotating subjects’ age could be summarized using three categories: (1)~to record the age of the subjects as provided by the subjects themselves, (2)~to infer subjects' age using other data sources, and~(3) to estimate subjects' age by observing their appearance. Of these methods, recording the subjects’ age-related information during an in-person data collection process (e.g. inviting participants to a studio and taking photos) was the most prevalent (58\%), though this often resulted in a smaller number of unique subjects represented in the data set with an average of 314.8 (std=386.5) people per data set.

Other methods for annotating subjects’ age provided a more scalable means to annotate ages. For instance, multiple data sets used publicly available dates for when a photo was taken and the subjects’ date of birth to infer their ages in that particular photo at a larger scale (21\%). This method was particularly common with data sets that contained images of celebrities as they appeared in movies or other public events as the date of the capture is given in the form of the release date of the movie, and as the date of birth of celebrities are easily accessible. This method provided the data sets with a relatively precise estimate of their subjects’ ages, although there could be some deviations as it usually takes some time for a movie to be released after it is filmed. A related (but less precise) method for annotating subjects’ age was to search for celebrities’ names on an online search engine, followed by a descriptor such as “young” and “old” to retrieve young and old-looking images of the same celebrity \cite{69_Bianco}. Such data sets that used other metadata to infer a subjects’ age included the Indian Movie Face Database and Cross-Age Celebrity data set and represented an average of 169,367.5 (std=406,955.0) people per data set.

Finally, some data sets employed crowdworkers on Amazon Mechanical Turk or students to study the appearance of the photos' subjects and manually annotate them based on how old the subjects looked (13\%). The training and annotation procedure for the annotators, if there was any, was not made clear in any of the data set documentation except for the categories with which the subjects were labeled. As covered above, there were no standards running through the categories used across these data sets. For instance, the CIFAR-100 data set labeled its subjects with weakly-defined categories such as “baby,” “boy,” “girl,” “man,” and “woman” \cite{68_Krizhevsky} while the 10K U.S. Adult Faces Database labeled them with unevenly spaced age categories that started from 20-30, 30-45, 45-60, and ended with over 60 years of age  \cite{66_Bainbridge}. These data sets contained an average of 5,384.1 (std=6,765.6) unique subjects but it is unclear how well the annotation reflects the ground-truth age of the subjects.\footnote{The average and standard deviation in this sentence was calculated without the value for the Real-World Affective Faces Database, which was only documented to include thousands of unique individuals without a precise value.}

\subsection{Goal of Gathering Age}
A total of 14 out of 24 data sets we studied that included age-related information did not specify why the subjects’ ages were collected (58\%), but rather noted in the documentation that the demographics of the subjects were included as a part of the data set distribution. For example, the documentation of CMU’s Multi-Pie Face Database writes: “As part of the distribution we make the following demographic information available: gender, year of birth, race and whether the subject wears glasses” \cite{72_Gross}. On the other hand, some data sets gave broad reasoning for including age metadata, suggesting that the goal of the data set is to provide a data set for face recognition or analysis tasks that covers a relatively diverse population (17\%) to serve as “an unbiased platform” for future studies \cite{66_Bainbridge}. 

However, of the 10 face data sets that more specifically described the goal of collecting the subjects’ age, the most common reasons for doing so were directly connected to supporting age-related classification or analysis tasks (60\%). For instance, the Iranian Face Database that contains face images of subjects between ages 2-85 was curated by the data set authors to support the creation of “a reliable age classification algorithm” that takes as an input a face image and outputs the age estimate of the subject in the image \cite{73_Bastanfard}. Meanwhile, data sets such as the Large Age-Gap data set \cite{69_Bianco} or Cross-Age Celebrity data set were created to provide longitudinal face data for a particular person to help create face recognition algorithms that can recognize a particular person at different ages, suggesting that such algorithms can help tag users on photo-sharing websites like Facebook and Flickr where users post images over many years. Of those data sets that explicitly covered the older adults category in their closed intervals for the subjects’ ages, a majority of them (60\%) were created in order to support such age-related classification or analysis tasks.

\subsection{Representation of Older Adults}
The U.S. census reported that 13.0\% of the adults in the U.S. are at least 65 years old while 1.9\% are at least 85 years old in 2010 \cite{AARP2010}. Both of these numbers are expected to grow significantly in the years to come, with the older adult population expected to account for 20.2\% and the oldest-old expected to account for 4.3\% of the U.S. population by 2050 \cite{AARP2010}. However, in the face data sets we studied, we found that the representation of older adults is not proportional to the population at large. 

This is particularly evident in the ranges for the subjects’ ages that 50\% of the data set documentation with age-related information provided. For example, the Long Distance Heterogeneous Face Database described the subjects’ age range in the data set as: “The 100 subjects who participated in our study (70 males and 30 females) were students at Korea University with an age range of 20-30 years old.” Among the data sets that provided such ranges in their documentation, we found that the average age of the oldest individuals included in the face data sets was 56.3 years old (std=19.3), which is lower than 65 years old, the lower bound for older adults that is commonly used in literature on age and technology \cite{Loi2021}. In addition, of those data sets whose documentation provided the age ranges, less than half of them included at least one subject who was older than 65 years old (42\%) with only the Iranian Face Database with images of 616 people including at least one subject who was older than 85 years old. However, even the Iranian Face Database was heavily skewed towards the younger generation; only 7.5\% of its subjects were included in the age categories greater than 60 years old and only 3.7\% in the age categories greater than 70, while its median age category was 21-30 years old. 

Other data set documentation summarized the subjects’ ages either with an open bracket (e.g. older than 50) (13\%) or by calculating the average and standard deviation of the subjects’ age (21\%).
Observing such ranges and aggregate statistics also presents a similar concern for under-representation of older adults. The highest age categories in those that summarized subjects’ age with an open bracket started well below the starting age of older adults; for example, the oldest category of “31 and above” used in GUC Light Field Face Artifact Database \cite{70_Raghavendra}. Meanwhile, the average age of the subjects that 21\% of 24 the documentation with age-related information provided averages to 25.00 years old (std=3.4). 

Overall, the 92 face data sets we studied are under-representing older adults while only one explicitly includes any represention at all of the oldest-old adults. But it is just as noteworthy that some data sets are also particularly skewed towards the younger population in their 20’s. Some of the documentation indicated that one possible explanation for this distribution is that the prevalence of younger adults may be an artifact of convenience sampling. 25\% of the data set documentation that included age-related information specifically mentions that the subjects were drawn from a university undergraduate population, for example as mentioned in the documentation of the NIMSTIM Set of Facial Expressions data set: “[the subjects] included… undergraduate students from a liberal arts college located in the Midwestern United States” \cite{71_Tottenham}.

\section{Discussion}
Ensuring that older adults are represented in the data sets used to train and/or test AI can help ensure that emerging AI tools will work well for this important and growing population. But creating more representative data sets for older adults should start from understanding how well they are represented right now. In this section, we synthesize what we learned from studying the 92 face data sets and discuss the extent of older adults’ representation. We provide concrete suggestions towards better representation of this population.  

\subsection{Older Adults Are Under-Represented}
That older adults are under-represented may not be surprising; they are often not the target user population for new technology, their data is less available on the web for scraping, and their data contributions may be less readily accessible to university researchers compared to convenience samples such as college students. But the extent to which they are under-represented is cause for concern. Less than half of the data sets whose documentation provided the maximum age of their subjects had at least one person older than 65 years old, while only one data set out of the 92 that we studied in this work explicitly had at least one person older than 85, the starting age for the fast-growing oldest-old adult category. On the other hand, younger people, particularly those in their undergraduate years, are much more heavily represented in the data sets, as the researchers who curate these data sets often recruit their subjects from the academic institutions that they are a part of. This skew towards the younger generation is highly reminiscent of many psychology studies that recruited from the undergraduate population, a field which now has growing concerns for the generalizability of research findings towards the larger population outside of a university \cite{74_Hanel, 75_Henrich}. 

\subsection{Challenges of Older Adults' Representation}
We note that there are unique challenges in creating representative data sets in terms of the subjects’ age when compared to other demographic categories. Firstly, the age distribution in the general population is fast-changing with the average lifespan of individuals increasing across different parts of the world \cite{77_UN, AARP2010}. It is possible that in the future, even the definition of the oldest-old may need to be revisited to include those who are much older than 85 years old (e.g. aged 100+). So although the data sets we studied continue to be used by the research community, their representativeness in terms of age will worsen as time advances. Further complicating matters, recent literature on aging suggests that as life expectancy increases, the way we age changes as well \cite{76_Jones}, making it potentially non-trivial to account for the current generation of older adults with the previous generation’s data even if varied age categories are represented.  

Intersectionality is also an important concern when creating age-representative data sets. For instance, are all the older adults in the data sets of certain genders or races? The older adults population is unevenly distributed across different countries, race, and gender identities \cite{77_UN, AARP2010}. Even if we make efforts to represent older adults in our data sets, if older adults are under-represented in certain intersectional demographics, we might repeat and propagate intersectional biases, an issue pointed out by the Gender Shades study that showed that AI systems'  performance disparity could be particularly severe for certain intersectional groups \cite{Buolamwini2018}. 

Given these challenges in ensuring representative AI data sets in terms of age, we see the core message of our work not only as one that suggests that older adults are dismissed when creating such data sets right now. Instead, our work also implies that a successful representation of older adults is a complex moving target for which we as a community need to make continuous efforts to understand the changing demographic makeup of our society and adapt. Thus we see our core contribution as not only determining whether we are succeeding in representing older adults but as starting a conversation about age representation that will inform future efforts to create more representative AI data sets. 

\subsection{Need For a Standardized Approach}
More concretely, what should we as a community of researchers and engineers creating future AI data sets strive towards? One important theme that arose from our results is the lack of a standardized approach for documenting age-related information. To start, we found that only about a quarter of the data sets we studied included some form of age-related information in their documentation (26\%), while a fifth did in their metadata (19\%). Although collecting such information could be challenging in certain scenarios, especially when the data is not directly sourced from the subjects and the ground-truth label is unattainable, the importance of making a conscious effort to create an age-representative data set needs to be highlighted. But beyond this, we found that the ways age was categorized were inconsistent across different data sets, with some using unevenly spaced age categories and some using more abstract categories like ``young'' and ``old.'' This makes interpreting and comparing the age representation across different data sets challenging. To remedy this, researchers could consider collecting the raw ages of their subjects if the data collection is taking place in-person. In cases where the raw ages are not available or collecting raw ages poses privacy concerns for the subjects, a possible source of inspiration we could draw from is to adhere to the age categories as they are presented in a large scale census (e.g. a government curated census), which would provide standardization as well as enabling comparisons regarding representation. If different categories of age are used, we suggest that they be motivated and defined. Finally, it should be noted that some methods for inferring a subject’s age (e.g. estimating age by observing the subject’s appearance) could reflect societal bias and other forms of inaccuracy. For instance, a recent work showed that age attribute is often labeled in a gender-dependant way and exhibited gender bias with crowd workers more likely to rate faces of men as “Young” and faces of women as ``Old'' \cite{Ramaswamy2020}. If such methods need to be employed to annotate the ages of subjects, we suggest that the annotation procedure be clearly documented.

\section{Limitations and Future Directions}
Our work presents a focused but poignant illustration of the under-representation of older adults in 92 face data sets used in the academic literature, and it is relevant to the greater discourse around AI, its ethics and fairness. However, we note important limitations to our work that suggest opportunities for future research. First, we focused on the representation of older adults in AI data sets but we did not directly cover the performance of the models trained with data sets that under-represent older adults. We see data representativeness as a worthwhile goal to pursue as what is used to train and test a model often directly correlates with the performance of that model when used by different demographics. Future studies can build on our findings by exploring how the under-representation of older adults translates to AI systems’ performance for this population in practice. A recent report from NIST \cite{Grother2019} suggests such performance disparities exist, noting that they observed an increase in false positives in face recognition among the older adults. 

Second, our exploration of face data sets should be considered as a case in point to illustrate the possible under-representation of older adults not only in face data sets but also in other forms of AI data sets. For instance, does the audio data for training voice recognition systems account for older adults with slower speech, and does the motion data for identifying moving pedestrians account for older adults with limited mobility? We hope our study functions to motivate future efforts to investigate and improve age representation in other types of AI data. Additionally, it is worth noting that our findings that suggest the uniquely challenging aspect of representing age in AI data sets, which is deeply transitory, resonate with the fluid manner in which people increasingly view other demographic categories such as gender and race. Subsequent work should continue to explore how to bridge this diverse and ever-developing demographic landscape with better representation in AI technology.

Finally, there are important normative questions around what representative training data sets and AI models would mean for older adults; what are the right use cases for the AI models that can be trained with data sets studied here, and how can we ensure that older adults actually reap the benefit from these systems? For instance, facial recognition technology has drawn concerns over the years that it might be vulnerable to abuse, especially when used in contexts such as automated surveillance \cite{Ruha2019}. In such cases, could better representation in AI data sets sometimes generate harm for the marginalized communities, and if so, how can we prevent or mitigate such harm? For the scope of this paper, we did not directly engage with these fundamental normative questions, but we believe that they should be continually discussed as we refine our AI models.

\section{Conclusion}
There will be meaningful opportunities in which future generations of AI systems may benefit older adults such as by maintaining physical independence and facilitating social connection, as well as interacting with the variety of AI-powered applications aimed toward the general public. Ensuring the success of such interaction may be dependent on whether the data sets that are created to train and test AI systems represent older adults. In this work, we explored 92 face data sets as a case study to investigate whether the age categories represented in these data sets reflect the fast-changing age distribution of the population at large. We highlight that older adults are under-represented in these data sets and that ensuring representation of various age demographics poses many challenges. Informed by our findings, we suggest more standardized practices for documenting and annotating subjects’ age in these data sets and call for our field’s continual efforts to curate representative and inclusive AI data sets, including with attention to age.

\bibliographystyle{ACM-Reference-Format}
\bibliography{main}

\end{document}